\def\imat {\rm i}

\documentstyle[twocolumn,floats,epsf,ifthen,aps,prb]{revtex}

\begin{document}

%\preprint{draft}

%\draft

\title{Entanglement of solid-state qubits by measurement}

\author{Rusko Ruskov\footnote{On leave from
INRNE, Sofia BG-1784, Bulgaria} 
and Alexander N. Korotkov} 
\address{
Department of Electrical Engineering, University of California,
Riverside, CA 92521-0204.
}
\date{\today}

\maketitle

\begin{abstract}
We show that two identical solid-state qubits can be made fully entangled 
(starting from completely mixed state) with probability 1/4 just
measuring them by a detector, equally coupled to the qubits. This
happens in the case of repeated strong (projective) measurements as well 
as in a more realistic case of weak continuous measurement. In the 
latter case the entangled state can be identified by a flat
spectrum of the detector shot noise, while the non-entangled
state (probability 3/4) leads to a spectral peak at the Rabi
frequency with the maximum peak-to-pedestal ratio of 32/3.
\end{abstract}
%\pacs{PACS numbers: }
%\pacs{03.65.Ud; 03.76.Lx; 73.23.-b} 
%PACS:  73.23.-b Electronic transport in mesoscopic systems
%       73.40.Gk Tunneling 
%       73.90.+f Other topics in electronic structure and electrical properties
%                of surfaces, interfaces, and low-dimensional structures 
%       03.65.Ud Entanglement and quantum nonlocality 
%       03.65.Yz Decoherence, open systems; quantum statistical methods
%       03.65.Ta Foundations of quantum mechanics; measurement theory 
%       03.67.Lx Quantum computation 

%\newpage
\narrowtext

%\vspace{1ex}
\vspace{0.6cm}

    Prospective solid-state realizations of quantum computers
\cite{Bennett} have many potential advantages over other proposed 
realizations, including more natural scalability, simple electrical
control of parameters, and use of very well developed technology. 
A number of theoretical proposals on solid-state quantum computers 
have been put forward (see, e.g.\ Refs.\ 
\cite{Kane,Loss-DiVincenzo,Mooij-th,Averin-FDP,Makhlin-RMP}) 
and spectacular experimental
results have been achieved, including demonstrations of charge qubits 
\cite{Nakamura} using single-Cooper-pair boxes, flux qubits 
\cite{Mooij,Lukens} using superconducting loops interrupted by Josephson 
junctions, and combined charge-flux qubits \cite{Vion} 
(the demonstrated quality factor 
was as high as 25,000 in Ref.\ \cite{Vion}). Obviously, next important 
experimental step is the demonstration of entangled solid-state qubits. 

        The qubits can be made entangled by their direct interaction if the 
interaction depends on the states of the qubits and at least one qubit
is in a coherent superposition before the interaction. In this letter
we will discuss an alternative way, when two solid-state qubits are made 
entangled just by their simultaneous measurement by one detector, which
thus provides an indirect coupling between qubits. The procedure works 
with a probability less than unity, and in this respect it is somewhat 
similar to the operation of conditional quantum gates 
\cite{Knill} based on simple linear optical elements. 
 Our procedure can also be thought of as an entanglement
purification \cite{Bennett-purif}, in which instead of trading the number 
of qubit pairs for better entanglement, we trade the success probability
while no initial entanglement is needed. 
     One more analogy is the preparation of entangled atoms in an optical 
cavity by monitoring the cavity decay \cite{Plenio}.

        In contrast to qubits represented by photons, which are physically
destroyed by the acts of measurement, solid-state qubits only change their
state due to measurement, that allows somewhat more freedom in designing 
quantum operations. On the other hand, it is quite difficult to realize
simple projective measurements of solid-state qubits because of typically 
weak coupling with detector. Therefore, instead of simple abrupt collapse,
we have to deal with dephasing-like processes in the case of ensemble
measurements \cite{Caldeira} or with the continuous (weak) 
measurements \cite{Davies,Wiseman,Kor-meas2,Goan} in the case of single 
qubits. 

        The theory of non-averaged continuous measurement of single 
solid-state 
qubits \cite{Kor-meas2,Goan} is under active development during last 4 years 
and has already produced a number of experimental predictions (for some of 
them, see Ref.\ \cite{Kor-last}). One of the predictions \cite{Kor-sp} 
is related to a direct continuous measurement of a qubit Rabi oscillations 
by a weakly coupled detector. The theoretical result is that the qubit 
oscillations can be evidenced by the peak of the detector current
spectral density at the Rabi frequency; however, the peak height cannot
be larger than 4 times the noise pedestal (this fact seems to have 
recent experimental confirmation \cite{Durkan}).

        In this letter we consider the case when two identical  
qubits perform Rabi oscillations, which are continuously measured by a 
detector equally coupled to two qubits. We will show that the system is 
gradually collapsed into one of the two regimes: 
either qubits become fully entangled (Bell 
state) that can be identified by absence of the spectral peak of the detector
current, or qubits state fall into the orthogonal subspace 
that can be identified by the Rabi spectral peak, 
which for an ideal detector is 32/3 times higher than the noise pedestal. 
The probabilities of two scenarios are 1/4 and 3/4, respectively (starting
from completely mixed state), so on average the peak-to-pedestal
ratio is equal to 8, twice as large as for a single qubit.

\begin{figure}
\centerline{
\epsfxsize=2.9in
\hspace{-0.4cm}
\epsfbox{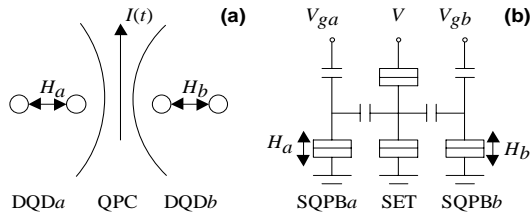}
%\epsfbox{qq.eps}
}
\vspace{0.2cm}
\caption{Schematic of two qubits measured by an equally coupled detector.
(a): Realization based on double quantum dots measured by a 
quantum point contact, (b): realization based on single-Cooper-pair-boxes
measured by a single-electron transistor. Measurement can entangle qubits. 
 }
\label{fig1}\end{figure}

        Figure \ref{fig1} shows possible realizations of our setup. 
In the first realization [Fig.\ \ref{fig1}(a)] each qubit 
is made of a double quantum dot \cite{Buks} (DQD), occupied by a single 
electron, while 
the detector  is a quantum point contact (QPC) located in between
DQDs. The second possible realization [Fig.\ \ref{fig1}(b)] is
based on single-Cooper-pair-boxes (SCPB) as qubits \cite{Nakamura}, which
are measured by a single-electron transistor (SET). Other possible
realizations (not shown) can be based on 
flux qubits \cite{Mooij,Lukens} or combined charge-flux qubits \cite{Vion}.

        In the Hamiltonian of the system,
\begin{equation}
  {\cal H} = {\cal H}_{QB} + {\cal H}_{DET} + {\cal H}_{INT} ,
\end{equation}
        the first term describes two qubits alone, 
${\cal H}_{QB}= (\varepsilon _a/2)(a_\downarrow^\dagger a_\downarrow 
         - a_\uparrow^\dagger a_\uparrow)
  +H_a (a_\uparrow^\dagger a_\downarrow + a_\downarrow^\dagger a_\uparrow)
+(\varepsilon _b/2)(b_\downarrow^\dagger b_\downarrow 
         - b_\uparrow^\dagger b_\uparrow)
+H_b (b_\uparrow^\dagger b_\downarrow + b_\downarrow^\dagger b_\uparrow)$,
where $\varepsilon_a$ and $\varepsilon_b$ are energy asymmetries, which
are neglected in this letter, $\varepsilon_a=\varepsilon_b=0$, the amplitudes 
$H_a$ and $H_b$ describe the tunneling within qubits (we assume $H_a=H_b$ 
in the most of this letter), and the direct interaction term  
$Ua_\uparrow^\dagger a_\uparrow b_\downarrow^\dagger b_\downarrow$ 
is neglected. 
The frequencies of free Rabi oscillations of qubits, 
$\Omega_a=(4H_a^2+\varepsilon_a^2)^{1/2}=2H_a$ and 
$\Omega_b=(4H_b^2+\varepsilon_b^2)^{1/2}=2H_b$ 
 (we use $\hbar=1$) 
obviously coincide, $\Omega_a=\Omega_b =\Omega$ 
[$\Omega \equiv (\Omega_a+\Omega_b)/2$] if $H_a=H_b$.  
For simplicity we limit ourselves by the case \cite{Gurvitz} 
of DQD qubits, measured by a low transparency QPC (though generalization
to other cases is simple), so that the detector Hamiltonian 
is ${\cal H}_{DET}=\sum_l E_l c_l^\dagger c_l + \sum_r E_r c_r^\dagger c_r
+\sum_{l,r} T(c_l^\dagger c_r + c_r^\dagger c_l)$ 
and the interaction term is 
${\cal H}_{INT}= \sum_{l,r} \Delta T_a (a_\uparrow^\dagger a_\uparrow 
         - a_\downarrow^\dagger a_\downarrow) 
  (c_l^\dagger c_r + c_r^\dagger c_l)
+\sum_{l,r} \Delta T_b (b_\uparrow^\dagger b_\uparrow 
         - b_\downarrow^\dagger b_\downarrow) 
  (c_l^\dagger c_r + c_r^\dagger c_l)$; we will be mostly interested in the
case of equal coupling, $\Delta T_a =\Delta T_b$. 

        The four basis states of two qubits, 
$|1\rangle \equiv \, \mid \uparrow_a \uparrow_b \rangle$, 
$|2\rangle \equiv \, \mid \uparrow_a \downarrow_b \rangle$, 
$|3\rangle \equiv \, \mid \downarrow_a \uparrow_b \rangle$, 
$|4\rangle \equiv \, \mid \downarrow_a \downarrow_b \rangle$, 
correspond to 4 values of the average current through the detector:
$I_{1,2,3,4} = 2\pi (T\pm \Delta T_a \pm \Delta T_b)^2 
\rho_l \rho_r e^2 V$,  
where $V$ is the QPC voltage and $\rho_{l(r)}$ are densities of states. 
It is important that in the case of equal coupling 
two currents coincide, $I_2=I_3\equiv I_{23}$, so the measurement
cannot distinguish between states $|2\rangle$ and $|3\rangle$. 
Besides the ``measurement'' basis, it is convenient to introduce also
the Bell basis: $|1\rangle^B \equiv (\mid \uparrow_a \downarrow_b \rangle 
  -\mid \downarrow_a \uparrow_b \rangle )/\sqrt{2}$,
$|2\rangle^B \equiv (\mid \uparrow_a \uparrow_b \rangle 
   -\mid \downarrow_a \downarrow_b \rangle )/\sqrt{2}$,
$|3\rangle^B \equiv (\mid \uparrow_a \downarrow_b \rangle 
   +\mid \downarrow_a \uparrow_b \rangle )/\sqrt{2}$, and
$|4\rangle^B \equiv (\mid \uparrow_a \uparrow_b \rangle 
   +\mid \downarrow_a \downarrow_b \rangle )/\sqrt{2}$. 
Notice that $|1\rangle^B$ and $|2\rangle^B$ are eigenstates 
of ${\cal H}_{QB}$ if $H_a=H_b$, while the states 
$|3\rangle^B$ and $|4\rangle^B$ are transformed by ${\cal H}_{QB}$ 
as $\cos (\Omega t+\phi )|3\rangle^B -\imat \sin (\Omega t+\phi)|4\rangle^B$.

        Before considering continuous measurements, 
let us discuss a simpler case of a sequence of ``orthodox'' projective 
measurements which can be realized if the coupling with the detector 
is strong (${\cal C} \gg 1$, see below) and detector voltage is applied 
during short time intervals. 
    Since the states $|2\rangle$ and $|3\rangle$ are mutually 
indistinguishable, 
the the two-qubit density matrix $\rho$ is projected each time 
into one of three subspaces, corresponding to states 
$|1\rangle$, $|23\rangle$, and 
$|4\rangle$ (we use notation $|23\rangle$ for the subspace spanned by
$|2\rangle$ and $|3\rangle$). The sequence of measurements separated
by time periods $\Delta t$ can be described by such projections separated
by intervals of unitary evolution due to ${\cal H}_{QB}$. 
 
Suppose the first measurement resulted in the current 
$I_{23}$, then the system is projected into $|23\rangle$ subspace, which is
also a subspace $|13\rangle^B$ in the Bell basis. If the state would be
exactly $|1\rangle^B$ (which does not evolve under ${\cal H}_{QB}$), then 
all subsequent measurements would give the same result $I_{23}$ and 
the state $|1\rangle^B$ would remain unchanged. However, if the two qubits 
would be in the state $|3\rangle^B$, then the next measurement would result
again in $I_{23}$ only with probability $p=(\cos \Omega \Delta t)^2$, while
the probabilities of results $I_1$ and $I_4$ would be $(1-p)/2$ each. 
Therefore, 
if a sufficiently long sequence of current measurements repeatedly gives the 
result $I_{23}$, the two-qubit density matrix $\rho$ purifies and 
becomes close to the fully entangled state $|1\rangle^B$. 

        Simple analysis shows that after $N$ successful
measurements (all results are $I_{23}$)
        \begin{equation}
\rho_{11}^B(N) =\frac{\rho_{11}^B (0)}{\rho_{11}^B (0)+
\rho_{33}^B (0) (\cos \Omega\Delta t)^{2(N-1)}} \, ,
        \label{rho(N)}\end{equation}
where $\rho_{11}^B (0)$, $\rho_{33}^B(0)$, and $\rho_{11}^B (N)$ 
are the corresponding density
matrix elements in the Bell basis before and after the sequence of 
measurements, while the probability of a successful sequence is 
$P(N)=\rho_{11}^B (0)+ \rho_{33}^B (0)(\cos \Omega\Delta t)^{2(N-1)}$. 
For large $N$ the difference between the obtained state and state 
$|1\rangle^B$ becomes exponentially small, 
while the probability of success 
is close to $\rho_{11}^B (0)$, which is equal to 1/4 for the fully mixed
initial state $\rho_{ij,mix}=\rho_{ij,mix}^B=\delta_{ij}/4$ (this state 
is a direct product of fully mixed states of each qubit).
        The purification rate depends on $\Delta t$,  
and is the fastest when $\Delta t$ is close to $(1/4+k/2)2\pi/\Omega$ ($k$ 
is an integer), which is a regime opposite to the quantum nondemolition
measurements \cite{Braginsky}.

        Notice that if some measurement in a sequence resulted in the 
current $I_1$ or $I_4$, then the subsequent measurements (for general 
$\Delta t$) can result in 
any current $I_1$, $I_4$, or $I_{23}$; however, $\rho_{11}^B$ remains
exactly zero (therefore, long sequences of $I_{23}$ results become extremely
improbable). Hence, to obtain the Bell state $|1\rangle^B$, 
one have to apply some perturbation which mixes two subspaces (for example,
apply a noise which affects $\varepsilon_a$ and/or $\varepsilon_b$) and 
repeat the procedure. Assuming a well-mixed case, the probability 
$1-(3/4)^{M}$ to obtain the state $|1\rangle^B$ can be made arbitrary close
to unity by allowing sufficiently large number $M$ of attempts.

        The procedure can obviously be used for the {\it preparation} 
of entangled states in a solid-state quantum computer, so it is important to 
discuss what happens if the conditions $H_a=H_b$ and $I_2=I_3$ are not 
satisfied exactly. In the case of slightly different $H_a$ and $H_b$,
Eq.\ (\ref{rho(N)}) changes insignificantly ($\cos \Omega \Delta t$ should
be replaced with $\cos \Omega\Delta t/\cos 2^{-1}\Delta\Omega\Delta t$, where
$\Delta\Omega\equiv \Omega_a-\Omega_b$), 
however, the probability of an $N$-long successful sequence becomes
$P(N)=\rho_{11}^B(0) (\cos 2^{-1}\Delta\Omega\Delta t)^{2(N-1)}+ 
\rho_{33}^B(0) (\cos\Omega\Delta t)^{2(N-1)}$ and decreases to zero
at $N\rightarrow\infty$. Estimating the average length of a successful 
sequence, $\overline{N}\simeq (\sin 2^{-1}\Delta \Omega\Delta t)^{-2}$, 
one can estimate a typical deviation from state $|1\rangle^B$,
        \begin{equation}
  1-\rho_{11}^B \sim (\cos\Omega \Delta t/\cos 2^{-1}\Delta\Omega\Delta t)^
{2/(\sin 2^{-1}\Delta \Omega \Delta t)^2} ,
        \label{rho(N)2}\end{equation}
which is as small as $\sim \exp[-(2\Omega/\Delta\Omega)^2]$ if 
$\Delta t\ll \Omega^{-1}$ (Quantum Zeno regime) and even smaller,
$\sim (\cos \Omega \Delta t)^{(32/\pi^2)(\Omega /\Delta \Omega)^2}$, 
if $\Delta t$ is close to $\pi/2\Omega$.

        To analyze the effect of a small difference between $I_2$ and $I_3$ 
because of slightly different coupling, we have to use the standard 
theory \cite{Davies} of weak quantum measurements and take into account 
the detector shot noise $S_i=2eI_i$. 
We assume that during
short measurement interval $\delta t$ the currents $I_1$ and $I_4$ 
can be unambiguously
identified, while current $I_2$ and $I_3$ are almost indistinguishable
and small corresponding signal-to-noise ratio is characterized by 
the parameter $\epsilon =(I_2-I_3)^2/4D \ll 1$ where $D=S_{23}/2\delta t$
is the variance of the measured noisy current. Each successful measurement 
tends to shift the state towards either $|2\rangle$ or $|3\rangle$ and
so decreases the amount of entangled state $|1\rangle^B$, that competes with
the purification due to Eq.\ (\ref{rho(N)}) and leads to iterative
formula $\rho_{11}^B(N+1)\simeq \rho_{11}^B(N)-\epsilon /4 
+[1-\rho_{11}^B(N)](\sin \Omega\Delta t)^2$ valid when $\rho_{11}^B$ 
is close to unity (exact formula is longer). Therefore, a typical deviation
from pure entanglement $1-\rho_{11}^B \simeq \epsilon/4(\sin \Omega\Delta
t)^2$ scales as $\epsilon$. 

        Now instead of instantaneous measurements let us consider 
a more realistic case of a continuous measurement, realized when 
the detector voltage 
is applied all the time. For the analysis we will use the Bayesian
formalism \cite{Kor-meas2} assuming weakly responding linear detecting 
regime, $|\Delta I_{a,b}|\ll I_i$,  $\Delta I_a\equiv I_1-I_3=I_2-I_4$, 
$\Delta I_b \equiv I_1-I_2=I_3-I_4$,  
and concentrating on the case of symmetric weak coupling, 
${\cal C}_a\approx {\cal C}_b \lesssim 1$, 
${\cal C}_a\equiv (\Delta I_a)^2/S_0H_a$,
${\cal C}_b\equiv (\Delta I_b)^2/S_0H_b$, where the frequency-independent 
detector shot noise spectral density $S_0$ does not depend significantly 
on the qubits state. 

        The evolution of the two-qubit density matrix $\rho$ can be
described by the equation \cite{Kor-ent} (in It\^o representation) 
        \begin{eqnarray}
\frac{d}{dt}\rho_{ij} = &&
( I(t)-\sum\nolimits_k \rho_{kk} I_k ) 
(I_i+I_j-2\sum\nolimits_k \rho_{kk}I_k) \, \frac{\rho_{ij}}{S_0}  
        \nonumber \\
&& -((I_i-I_j)^2/4S_0 +\gamma_{ij})\, \rho_{ij} 
-\imat [{\cal H}_{QB}, \rho]_{ij} ,
        \label{Bayes1}\end{eqnarray}
where the extra dephasing rate $\gamma_{ij}=(\eta^{-1}-1)(I_i-I_j)^2/4S_0$ 
depends on detector ideality $\eta$ ($0\leq \eta \leq 1$) and 
vanishes for the QPC as a detector\cite{Kor-meas2} ($\eta =1$); 
however, this term is important, for example, for the SET. To simulate
individual realizations of the random measurement process, the noisy
detector current $I(t)$ can be calculated as 
        \begin{equation}
I(t) =\xi(t) + \sum\nolimits_k \rho_{kk} I_k, 
        \label{Bayes2}\end{equation}
where $\xi (t)$ is a white noise with spectral density $S_0$. 
Notice that the averaged dynamics (master equation) can be obtained
by averaging over noise $\xi (t)$ that eliminates the first term 
in Eq.\ (\ref{Bayes1}).

\begin{figure}
\centerline{
\epsfxsize=2.9in
\hspace{-0.4cm}
\epsfbox{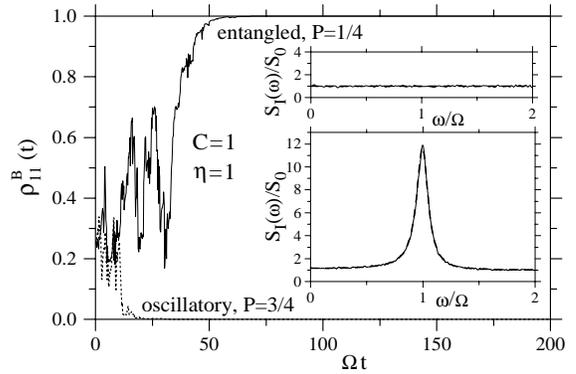}
}
\vspace{0.2cm}
\caption{ Two numerical realizations of $\rho_{11}^B$ evolution starting 
from the fully mixed state. The upper (solid) line illustrates the scenario 
of collapse into fully entangled Bell state $|1\rangle^B$, while the lower 
(dotted) line  
shows a collapse into the orthogonal subspace. Two insets show the 
corresponding spectral densities $S_I(\omega )$ of the detector noise 
(solid/dashed lines are the numerical/analytical resuls).
 }
\label{fig2}\end{figure}

        We have performed extensive Monte Carlo simulations \cite{Kor-meas2}
and found the following (Fig.\ \ref{fig2}). 
In the symmetric case, $H_a=H_b$, ${\cal C}_a={\cal C}_b={\cal C}$ 
[${\cal C}\equiv ({\cal C}_a+{\cal C}_b)/2$, we mostly used ${\cal C}$ 
between 1/4 and 1], any initial state 
either evolves eventually into the fully entangled Bell state 
$|1\rangle^B$ ($\rho_{11}^B \rightarrow 1$) or ends up in the 
orthogonal subspace ($\rho_{11}^B\rightarrow 0$) performing oscillations
within this subspace so that the ``signal'' $z\equiv \rho_{11}-\rho_{44}=
 2 \mbox{Re} \rho_{24}^B$ (which affects the detector 
current) oscillates with frequency $\Omega$ and fluctuating  amplitude 
within the range from 0 to 1. Within the accuracy consistent with the number 
of trials, the probability of evolving into state $|1\rangle^B$ coincides
with $\rho_{11}^B(0)$, similar to the case of orthodox measurement sequence.
For $\eta =1$ the two-qubit state  eventually becomes pure 
independently of an initial choice (similar to the one-qubit case 
\cite{Kor-meas2,Kor-last}).  
This fact is obvious for the state $|1\rangle^B$, while in the oscillating 
scenario the surviving nondiagonal matrix elements in the Bell basis 
satisfy equations 
$(\mbox{Re}\rho_{24}^B)^2=\rho_{22}^B\rho_{44}^B$, 
$(\mbox{Im}\rho_{23}^B)^2=\rho_{22}^B\rho_{33}^B$, 
and $(\mbox{Im}\rho_{34}^B)^2=\rho_{33}^B\rho_{44}^B$.  
In the state $|1\rangle^B$ the numerically calculated spectral
density of the detector current is flat and equal to $S_0$ 
(the signal $z$ is zero), while   
in the oscillating state it exhibits a peak (lower inset in Fig.\ \ref{fig2})
at frequency $\Omega$ 
with the peak height close to the analytical result $(32/3)S_0$ (see below).

        The fact of collapsing eventually either into the state $|1\rangle^B$
or into the orthogonal subspace can be understood using an analogy with
the sequential measurement case considered above, and is caused by the fact
that neither unitary evolution due to ${\cal H}_{QB}$ nor nonunitary
evolution due to measurement mixes two subspaces [see Eq.\ (\ref{Bayes1})]. 
The probability of two
scenarios should be equal to the contribution of two subspaces
to the initial density matrix, $\rho_{11}^B(0)$ and $1-\rho_{11}^B(0)$, 
since ensemble averaged value $\langle \rho_{11}^B (t) \rangle$ does  not 
change with time (as follows from the master equation). 

        To find analytically the spectral density of the detector current
for the oscillating state, we have used two 
approaches \cite{Kor-sp} 
leading to the same result. The first one is based on the master equation
and collapse ansatz. Using classical equation $I(t)=z\Delta I +\xi (t)$, 
we calculate the current correlation function 
$K_I(\tau )=\langle I(0) I(\tau )\rangle$ as 
$K_I(\tau >0 )=(\Delta I)^2 K_{\hat z}(\tau )$,
while $K_{\hat z}(\tau )$ is calculated in the following way.
At time $\tau =0$ the two-qubit state is collapsed into one of the three 
basis states of the subspace: $|1\rangle$ (corresponding to $I_1$ and $z=1$),
$|4\rangle$ (corresponding to $I_4$ and $z=-1$), or
$|3\rangle^B$ (corresponding to $I_{23}$ and $z=0$). The probabilities
of these collapses are 1/3 each, since for $\rho_{11}^B=0$ 
the stationary solution of 
the master equation is $\rho_{11}=\rho_{44}=\rho_{33}^B=1/3$  
(the stationary solution is diagonal in the Bell basis as well as in the basis
$\{|1\rangle,|4\rangle,|3\rangle^B\}$ and has zero entanglement 
\cite{Wooters}). In each of 3 cases, the value of $z$ at 
time $\tau$ is obtained from the solution of the master equation for 
two relevant components: 
	\begin{equation}
dz/d\tau = -\Omega y, \,\,\,\, dy/d\tau = \Omega z -\Gamma y, 
	\end{equation}
where $y\equiv 2 \mbox{Im} \rho_{23}^B$ and 
$\Gamma =\eta^{-1}(\Delta I)^2/4S_0$. 
So, $z(\tau ) =\pm G(\tau )$, 
$G(\tau )\equiv \exp (-\Gamma \tau /2) 
[\cos \tilde\Omega \tau +(\Gamma /2\tilde\Omega)\sin \tilde\Omega \tau]$ 
[here $\tilde\Omega \equiv (\Omega^2-\Gamma^2/4)^{1/2}$] in the 
two first cases while $z(\tau )=0$ in the third one. 
Summing 3 contributions to $\langle z(0)z(\tau )\rangle$ with probability 
weights 1/3 each, we obtain $K_{\hat z} = (2/3) G(\tau )$ 
and the current spectral density 
        \begin{equation}
    S_I(\omega ) = S_0 + \frac{8}{3}\, \frac{\Omega^2(\Delta I)^2\Gamma}
{(\omega^2-\Omega^2)^2+\Gamma^2\omega^2}. 
        \label{spectrum}\end{equation}
In the case $\Gamma \ll \Omega$ the spectral peak at the Rabi 
frequency $\Omega$  corresponds to the $Q$-factor of 
$8\eta /{\cal C}$ (similar to one-qubit case \cite{Kor-sp}) and has
the peak height equal
to $(32/3)\eta S_0$, confirming the numerical result (see lower inset
in Fig.\ \ref{fig2}). 

        The second method of $S_I(\omega )$ calculation is based on 
the Bayesian Eq.\ (\ref{Bayes1}) assuming $\eta=1$ and random evolution of
a pure state with 
$z=A(t)\cos (\Omega t +\Phi (t))$ [then $y = 
A(t)\sin (\Omega t+\Phi (t))$]. In this method \cite{Kor-sp} the 
correlation between
noise $\xi (0)$ and evolution of the density matrix at later time should be
taken into account, so $K_I(\tau >0)=(\Delta I)^2 K_z(\tau )+
\Delta I K_{\xi z}(\tau )$, while correlation functions $K_z(\tau )$ and
$K_{\xi z}(\tau )$ should be calculated by averaging of a long individual 
realization over time. We have proved \cite{next} 
that the result for $K_I(\tau )$ 
calculated by this method coincides with the result of the previous method
for arbitrary coupling ${\cal C}$; 
however, the formalism is much simpler for weak coupling, ${\cal C}\ll 1$. 
In this case the stochastic differential equations for $A(t)$ and $\Phi (t)$
can be averaged over oscillations with frequency $\Omega$ and the 
correlation functions can be calculated analytically \cite{next}: 
$K_z (\tau )= (5/12)G(\tau )$ and $K_{\xi z}(\tau >0)= G(\tau )\Delta I/4$. 
This gives us a natural
partition of the relative spectral peak height 32/3 into two contributions:
``classical'' part 20/3 comes from oscillations \cite{max} of the signal $z$, 
while the ``quantum'' contribution equal to 4 is due to
partial collapse of $\rho$ correlated with the 
detector noise. Comparing 
this partition with the partition $4=2+2$ for a one-qubit measurement 
\cite{Kor-sp}, we see that the classical part grows faster than the quantum 
part when the number of qubits is increased.

    Numerical simulations show that if the two Rabi
frequencies $\Omega_a$ and $\Omega_b$ are slightly different, or 
small difference between ${\cal C}_a$ and ${\cal C}_b$ is due to
asymmetry of coupling (different $\Delta I_a$ and $\Delta I_b$),
than the two-qubit density matrix $\rho$ 
makes rare abrupt jumps between a state very close to $|1\rangle^B$ 
and the oscillating state. 
 To find the switching rate analytically, 
we have used the master equation starting from entangled initial 
condition $\rho_{11}^B=1$ and calculated the linear term in $\rho_{11}^B (t)$
dependence at $t\gg \Gamma^{-1}$ (but when $\rho_{11}^B(t)$ is still close
to unity). In this way \cite{next} we have obtained the rate 
$\Gamma_{B\rightarrow O}=
(\Delta \Omega)^2/2\Gamma$ of switching from the Bell state to the
oscillating state due to slightly different Rabi frequencies, and the rate 
$\Gamma_{B\rightarrow O}=(\Delta {\cal C}/{\cal C})^2\Gamma /8$ when
$\Omega_a=\Omega_b$, but couplings $\Delta I_a$ and $\Delta I_b$ are
slightly different. To find the rate of the reverse switching, notice
that the stationary master equation has the solution $\rho_{ij,st}^B=
\rho_{ij,st}=\delta_{ij}/4$ (two subspaces are mixed now), 
therefore the system should spend on the average 1/4 
of the time in the state $|1\rangle^B$, and so 
$\Gamma_{O\rightarrow B} = \Gamma_{B\rightarrow O} /3$. 
The numerical histograms of switching time distributions confirm these
formulas. 
	Taking into account rare switching events, the average spectral 
density of the detector current is given by Eq.\ (\ref{spectrum}) multiplied
by 3/4, so the spectral peak height is equal to $8\eta S_0$.

        Finally, we have studied the effect of environmental dephasing,
modeling it with two small dephasing rates $\gamma_a$ and $\gamma_b$ acting
separately onto two qubits. This leads to a slightly mixed $\rho$ even for
an ideal detector and to switching events \cite{next} (similar to the 
case above) with $\Gamma_{B\rightarrow O}=3\Gamma_{O\rightarrow B}=
(\gamma_a+\gamma_b)/2$. Notice that a controllable weak external noise 
can be used in a simple feedback protocol to restore the entangled state 
after undesirable switching to the oscillating state. 

        In conclusion, we have found that the continuous measurement of two
identical solid-state qubits by equally coupled detector leads 
to either full spontaneous entanglement of qubits (Bell state) 
or to collapse into orthogonal oscillating state. 
In the latter case the noisy detector 
current has a spectral peak at the Rabi frequency, while in the former 
case the spectrum is flat. Slight asymmetry of the two qubit configuration 
as well as environmental dephasing leads to switching 
between two regimes. It is important to mention that 
for an experimental  observation  of  the  phenomenon  the  quantum 
ideality $\eta$
of the detector should not necessarily be close to unity; it should only
be large enough to allow distinguishing the Rabi spectral peak 
with the peak-to-pedestal ratio of $32\eta /3$.

The work was supported by NSA and ARDA under ARO grant DAAD19-01-1-0491.

%\vspace{-0.3cm}


\begin{references}
\bibitem{max} Notice that if $z$ would oscillate with the maximum possible
      amplitude of 1, the corresponding contribution to the peak 
	height would be equal to $8S_0$, which is still less than $32S_0/3$.
\end{references}
\end{document}